\documentclass[aps,twocolumn, superscriptaddress]{revtex4-1}
\usepackage[pdftex]{graphicx}% Include figure files
 \usepackage{amsmath}
\usepackage[T1]{fontenc}
\usepackage{amsfonts}
\usepackage{lipsum}
\usepackage{amssymb, mathrsfs}
\usepackage{braket}
\usepackage{subfigure}
\def\beq{\begin{equation}}
\def\eeq{\end{equation}}
\def\bsp{\begin{split}}
\def\esp{\end{split}}
\def\bea{\begin{eqnarray}}
\def\eea{\end{eqnarray}}
\def\ba{\begin{array}}
\def\ea{\end{array}}

\def\dg{\dagger}

\def\lb{\left(}
\def\rb{\right)}

\def\l.{\left.}
\def\r.{\right.}

\def\ra{\rangle}
\def\la{\langle}

\def\bo{\bold{k}}

\begin{document}

\title{Floquet Topological Magnons}
\author{S. A. Owerre}
\affiliation{Perimeter Institute for Theoretical Physics, 31 Caroline St. N., Waterloo, Ontario N2L 2Y5, Canada.}

\begin{abstract}
We introduce the concept of Floquet topological magnons --- a mechanism by which a synthetic tunable Dzyaloshinskii-Moriya interaction  (DMI)  can be generated in quantum magnets using circularly polarized electric (laser) field.  The resulting effect is that Dirac magnons and nodal-line magnons in  two-dimensional (2D) and  three-dimensional (3D) quantum magnets can be tuned to magnon Chern insulators and Weyl magnons respectively under  circularly polarized laser field.   The Floquet formalism also yields a tunable intrinsic DMI in insulating quantum magnets without an inversion center. We demonstrate that the Floquet topological magnons possess a finite thermal Hall conductivity  tunable  by the  laser field. 
\end{abstract}
\maketitle

Quantum magnets without an inversion center allow an intrinsic Dzyaloshinskii-Moriya    interaction (DMI) \cite{dm,dm2}, which is a consequence of spin-orbit coupling  (SOC) and it is usually fixed in different magnets.   The associated magnon bands in the magnetically ordered systems have a nontrivial topology with Chern number protected chiral magnon edge modes in 2D systems and magnon surface states in 3D systems. They are dubbed topological magnon Chern insulators \cite{lifa, alex4,shi,sol,kkim,sol1,xc}  and Weyl magnons \cite{fei,mok,su,su1} respectively. They are the analogs of electronic topological (Chern) insulators \cite{top3,top4, fdm} and Weyl semimetals \cite{xwan,lba,aab}. However, due to the charge-neutral property of magnons, topological magnonic materials are  believed to be potential candidates for low-dissipation magnon transports in insulating quantum magnets and they are applicable to  magnon spintronics  and magnetic data storage \cite{magn}. Topological magnonic materials ({\it i.e.}~magnon Chern insulators and Weyl magnons) also possess a thermal Hall effect  as  predicted theoretically \cite{alex0,alex5,alex2, alex22,shin} and observed experimentally  \cite{alex1,alex6, alex1a}. To date, topological magnon bands have been realized only in a quasi-2D kagom\'e ferromagnet \cite{alex5a}. 

Generally, every quantum  ferromagnetic material does not have a strong intrinsic  DMI necessary for topological magnons to exist.  For instance, the  single crystals   of the ferromagnetic  honeycomb  compounds CrX$_3$ (X $\equiv$ Br, Cl, and I)  show no evidence of  DMI \cite{dav0,dav,dav1,foot,foot1,foot2}, and kagom\'e  haydeeite also does not have  a finite (topological) energy gap in the observed spin-wave spectra \cite{bol},  suggesting that the DMI does not play a significant role in haydeeite.  These 2D ferromagnetic  materials with negligible DMI are candidates for Dirac magnons \cite{frans}, and  the 3D ferromagnetic counterparts  are candidates for nodal-line magnons \cite{mok1,su1}. By applying a circularly polarized laser field in these ``topologically trivial'' systems, one can generate topological magnons ({\it i.e.},~magnon Chern insulators and Weyl magnons) via a tunable synthetic laser-induced DMI. This is particularly important as it offers a way to tune the  DMI in magnetic materials. 

The Floquet theory \cite{shir} of laser-driven systems provides a theoretical as well as experimental method to engineer such  topologically nontrivial  systems. This formalism is mostly dominated by electronic systems \cite{foot3,foot4,foot5,gru,del,fot,fot1,jot,fla,we1,we2,we3,we4,we5,we6, gol,buk,eck,eck1,ste} and also optical bosonic  systems \cite{ple,stru,stru1,stru2,kenn,ew}. Moreover, an applied laser  field   provides a means for coherent control of the magnetization  in 1D quantum magnets  \cite{tak,tak1,tak2, taks, takh}.

In this Letter,  we introduce the Floquet formalism to 2D and 3D ferromagnetic quantum magnets in the presence of a circularly polarized electric (laser)  field. We show that the underlying charge-neutral magnons acquire a time-dependent Aharonov-Casher phase \cite{aha}, which generates a tunable synthetic DMI in magnetic systems with  negligible  intrinsic DMI. This leads to emergent nontrivial magnon Chern insulators and Weyl magnons with finite thermal Hall conductivity, which can be manipulated by the laser field.  Our results provide a novel platform to engineer    topological magnons in topologically trivial insulating quantum magnets. We hope that these results will extend the experimental search for topological magnons to a broader class of insulating quantum magnetic materials even without an  intrinsic DMI.  

Let us consider   2D and 3D ferromagnetic spin systems   described by the pristine Hamiltonian \begin{align}
\mathcal H_0&=  -J\sum_{\la \alpha,\beta\ra }{\bf S}_{\alpha}\cdot{\bf S}_{\beta},
\label{h}
\end{align}
where $\la \alpha,\beta\ra$ denotes summation over nearest-neighbour (NN)  sites and ${\bf S}_\alpha$ are the magnetic spin vectors at the lattice sites $\alpha$ located at ${\bf r}_\alpha$ and $J$ is the ferromagnetic interaction. For now we take the intrinsic DMI to be negligible, which is the case in most ferromagnetic compounds. Therefore, the goal is to generate a synthetic DMI by periodic modulation of the lattice using a time-dependent electric field of a laser  light.

We are interested in the underlying  magnon excitations of the quantum spin Hamiltonian \eqref{h} as described by the Holstein-Primakoff transformation:  $S_{\alpha}^{ z}= S-a_{\alpha}^\dagger a_{\alpha},~S_{\alpha}^+\approx \sqrt{2S}a_{\alpha}=(S_{\alpha}^-)^\dg$, where $a_{\alpha}^\dagger (a_{\alpha})$ are the bosonic creation (annihilation) operators, and  $S^\pm_{\alpha}= S^x_{\alpha} \pm i S^y_{\alpha}$ denote the spin creation and annihilation  operators which correspond to the hopping terms.  However, we will retain the spin operators in order to show the explicit form of the laser-induced DMI. As magnons are charge-neutral bosonic quasi-particles they do not interact with electromagnetic field except through their magnetic dipole moment, which we assume to be in the $z$-direction $\boldsymbol{ \mu}=g\mu_B \bold{\hat{z}}$, where $\mu_B$ is the Bohr magneton and $g$ the Land\'e $g$-factor. Now suppose that  a circularly polarized laser (electromagnetic) field with dominant oscillating electric field components  $\bold{E}(t)$ is irradiated perpendicular to the magnetic material lying on the $x$-$y$ plane. The magnetic dipole moment of magnon quasi-particles hopping in such an electric field background will acquire the   time-dependent Aharonov-Casher phase \cite{aha}
\begin{align}
\theta_{\alpha\beta}(t)=\frac{g\mu_B}{\hbar c^2}\int_{\bold{r}_\alpha}^{\bold{r}_\beta} \big(\bold E(t)\times \hat z\big)\cdot d\boldsymbol{\ell},
\label{pha}
\end{align}
where $\hbar$ is the reduced Planck's constant and $c$ is the speed of light. 
 The oscillating electric field obeys the relation $\bold E(t)=-\partial{\bold A(t)}/\partial t$, where ${\bf A}(t)$ is the time-periodic vector potential. We choose an  oscillating electric field   such that 
 \begin{align}
 {\bf E}(t) \times \hat z=E_0(\sin\omega t, -\cos\omega t, 0),
 \end{align}
where $E_0$ is the amplitude of the electric field. The corresponding time-dependent  Hamiltonian is given by

\begin{align}
\mathcal H(t)=-\sum_{\la\alpha,\beta\ra}\bigg[ \frac{J}{2}\lb S_\alpha^- S_\beta^+ e^{i\theta_{\alpha\beta}(t)}+ h.c.\rb + JS_\alpha^zS_\beta^z\bigg],
\label{flo2}
\end{align}
 Noticing that the direction of the vector pointing from $\bold{r}_\alpha$ to $\bold{r}_\beta$ defines a relative angle $\phi_{\alpha\beta}$, we get  $\theta_{\alpha\beta}(t)=\lambda\sin(\omega t-\phi_{\alpha\beta})$ where 
\begin{align}
\lambda = \frac{g\mu_B E_0 a}{\hbar c^2},
\label{dims}
\end{align}
is the dimensionless parameter characterizing the light intensity in the magnonic Floquet formalism and $a$ is the lattice constant. Notice the difference between the dimensionless parameter in Eq.~\eqref{dims} and that of  electronic Floquet formalism  \cite{foot4, foot5}.

The Floquet theory is a standard mechanism to study driven quantum systems in the likes of Eq.~\ref{flo2} \cite{foot3,foot4,foot5,fot}. It enables one to  transform a time-dependent model into a static effective model governed by what is called the Floquet Hamiltonian.  Now, we proceed with this formalism.    The static time-independent effective Hamiltonian can be expanded in the power series of $\omega^{-1}$ and it is given by $
\mathcal H_{\text{eff}}=\sum_{i\geq 0} \omega^{-i}\mathcal H_{\text{eff}}^{(i)}.$ We calculate the series expansion of the effective Hamiltonian using  the discrete Fourier component of the time-dependent Hamiltonian $\mathcal H^n=\frac{1}{T}\int_0^T dt e^{-in\omega t} \mathcal H(t)$ with period  $T=2\pi/\omega$, where \begin{align}
\mathcal H^n&=-\sum_{\la\alpha\beta\ra}\Big[\frac{J_{n,\perp}}{2}\lb S_\alpha^- S_\beta^+ e^{-in\phi_{\alpha\beta}}+ h.c.\rb + J\delta_{n,0}S_\alpha^zS_\beta^z\Big],
\label{flo1}
\end{align}
with $J_{n,\perp}=J\mathcal J_n(\lambda)$, and $\mathcal J_n(\lambda)$ is the Bessel function of order $n\in \mathbb Z$.   $\delta_{n,\ell}=1$ for $n=\ell$ and zero otherwise. We have used the standard  relation $e^{[iz\sin(x)]}=\sum_{m=-\infty}^{\infty}\mathcal J_m(z)e^{imx}$.
In the high frequency limit $\omega\gg J$ the leading order contribution is the zeroth order effective Hamiltonian given by $
 \mathcal H_{\text{eff}}^{(0)}=\mathcal H^0$,
where
\begin{align}
 \mathcal H_{\text{eff}}^{(0)}&=-\sum_{\la\alpha\beta\ra}\big[ J_{0,\perp} (S_\alpha^x S_\beta^x+ S_\alpha^y S_\beta^y) + J S_\alpha^zS_\beta^z\big],
 \label{zeroth}
\end{align}
and $J_{0,\perp}=J\mathcal J_0(\lambda)$. Thus, the zeroth order term yields an XXZ ferromagnetic Hamiltonian. For quantum spin-$1/2$ ferromagnetic systems, Eq.~\ref{zeroth}  is equivalent to the Bose-Einstein condensation of hardcore bosons studied by Matsubara and Matsuda  \cite{mat},  but in this case $J\geq |J_{0,\perp}|$. By lowering the frequency the first order contribution to the effective Hamiltonian is non-negligible.  It is given by
\begin{align}
\mathcal H_{\text{eff}}^{(1)}=\sum_{n=1}^{\infty}\frac{1}{n}\big[\mathcal H^n, \mathcal H^{-n}\big].
\end{align}
The commutator of the spin operators results in a product of three spins reminiscent of the scalar spin chirality. In order to see this we note that the $z$-component of the spins vanishes for $n\geq1$,  therefore the commutation relation in the first order term involves only  $\big[ S_\alpha^+ S_\beta^-, S_\rho^+ S_{\gamma}^-\big]=2\lb \delta_{\beta\rho}S_{\beta}^zS_\alpha^+ S_\gamma^- - \delta_{\alpha\gamma}S_{\alpha}^zS_\rho^+ S_\beta^-\rb$. We use this relation together with the identity  $\mathcal J_{-n}(z)=(-)^{n}\mathcal J_n(z)$ and obtain
\begin{align}
 \mathcal H_{\text{eff}}^{(1)}=\sum_{\Delta/\nabla} J_{\alpha\beta}^{(1)} {\bf S}_\gamma\cdot({\bf S}_\alpha\times {\bf S}_\beta),
 \label{syn}
\end{align}
where \bea J_{\alpha\beta}^{(1)}=-\sum_{n=1}^\infty 2 (-)^nJ_{n,\perp}^2\sin\lb n \Phi_{\alpha\beta}\rb,\eea  with $\Phi_{\alpha\beta}=\phi_{\alpha\gamma}-\phi_{\beta\gamma}$. ${\bf S}_\gamma=S_\gamma^z {\bf \hat z}$ and  $\gamma$ is the intermediate lattice site between $\alpha$ and $\beta$ \cite{alex0} and the sum is over the triangular plaquettes of the chosen lattice geometry.

The main result of this Letter is the induced  synthetic DMI by circularly polarized laser field as given in Eq.~\eqref{syn}. The  DMI  points along the  $z$-direction perpendicular to the $x$-$y$ plane. On the kagom\'e and pyrochlore lattices the synthetic DMI lies within the bonds of the NN sites, whereas on the honeycomb lattice it lies within the bonds of the next-nearest neighbour (NNN) sites. The DMI is the primary source of magnon Chern insulators and Weyl magnons in insulating quantum ferromagnets.
\begin{figure}
\includegraphics[width=2in]{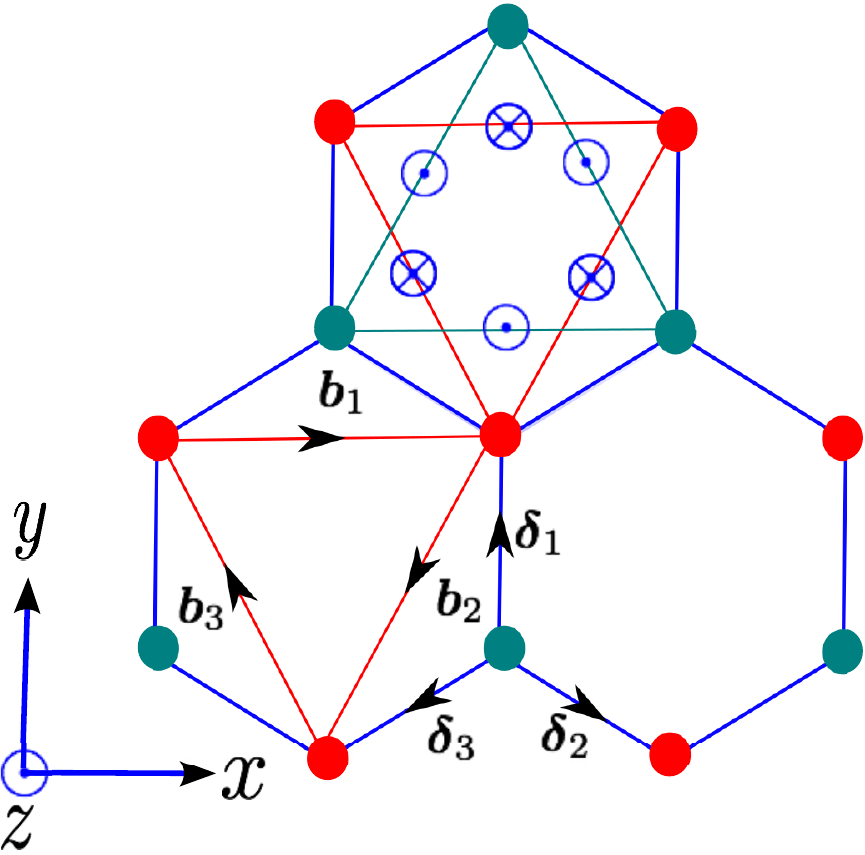}
\caption{The schematic of the honeycomb lattice with laser-field-induced DMI. The blue dotted and crossed circles denote the direction of the synthetic DM vectors pointing in and out of the lattice plane on the NNN bonds. Here, $\boldsymbol{\delta}_i$ are the vectors connecting the NN sites and $\boldsymbol{b}_i$ connect the NNN sites. The different  colours of the solid circles red(green) are used to denote the two sublattices $A$($B$).}
\label{lat}
\end{figure}
 Consequently,  Dirac magnons  in 2D ferromagnetic systems and nodal-line magnons  in 3D ferromagnetic systems with negligible intrinsic DMI  will be driven to   magnon Chern insulators and  Weyl magnons respectively  by the circularly polarized laser field. 
 
Now, we exemplify this method  by utilizing the 2D honeycomb ferromagnets for simplicity (see Fig.~\ref{lat}). 
In  this case the relative angle is given by  $\phi_{\alpha\gamma}-\phi_{\beta\gamma}=\frac{2\pi}{3}\nu_{\alpha\beta}$ \cite{eck,eck1}, where $\nu_{\alpha\beta}=+/-$ for hopping within $\Delta/\nabla$ respectively. Hence 
$J_{\alpha\beta}^{(1)}\approx \sqrt{3}\nu_{\alpha\beta}J_{1,\perp}^2.
$
\begin{figure}
\includegraphics[width=3.2in]{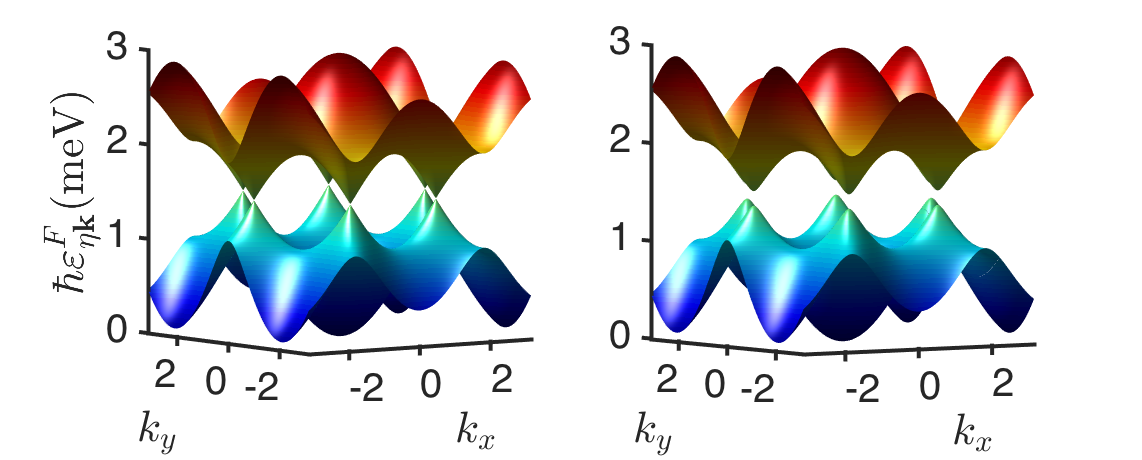}
\caption{Floquet magnon band structure of circularly laser-driven honeycomb quantum ferromagnetic spin system for $J=1$~meV $ ~\lambda=1.5$ in units of $g\mu_Ba/\hbar c^2$ and $\hbar\omega=2.5$. $(i)$ Floquet Dirac magnon in the zeroth order approximation $D_F=0$. $(ii)$ Floquet topological magnon in  the first order approximation  $D_F\simeq 0.014J$.}
\label{band}
\end{figure}
The total effective Floquet Hamiltonian up to first order is given by
\begin{align}
 \mathcal H_{\text{eff}}&=-\sum_{\la\alpha\beta\ra}\big[ J_{0,\perp} (S_\alpha^x S_\beta^x+ S_\alpha^y S_\beta^y) + J S_\alpha^zS_\beta^z\big] \label{mod}\\&\nonumber+D_F\sum_{\Delta/\nabla}\nu_{\alpha\beta}  {\bf S}_\gamma\cdot({\bf S}_\alpha\times {\bf S}_\beta),
\end{align}
where $D_F=\sqrt{3}J_{1,\perp}^2/\omega$. We note that $|J_{0,\perp}|\leq J$, so that the spins would prefer to order magnetically along the $z$-direction. The synthetic DMI does not affect the ferromagnetic ordering of the spins.  We study the underlying magnon excitations of Eq.~\ref{mod} by implementing the linear order Holstein-Primakoff transformation mentioned above on the two sublattices $A$ and $B$ of the honeycomb lattice.   In the bosonic representation the synthetic DMI is imaginary. It generates a synthetic magnetic gauge flux within the NNN sites.  The momentum space Hamiltonian is given by
\begin{align}
 \mathcal H_{\text{eff}}(\bo)=h_0{\bf I}_{2\times 2} +\bold{h}(\bo)\cdot\boldsymbol{\sigma},
\end{align}
where $\boldsymbol{\sigma}=(\sigma_x, \sigma_y, \sigma_z)$ are Pauli matrices and ${\bf I}_{2\times 2}$ is an identity $2\times 2$ matrix. $h_0= 3JS$, and $\bold{h}(\bo)=(h_x(\bo),h_y(\bo),h_z(\bo))$, with  $h_x(\bo)=-J_{0,\perp} S\sum_j\cos \bo\cdot\boldsymbol{\delta}_j$, $h_y(\bo)=-J_{0,\perp} S\sum_j\sin \bo\cdot\boldsymbol{\delta}_j$, and $h_z(\bo)=2D_F S^2 \sum_j\sin \bo\cdot\boldsymbol{b}_j$. The  Floquet magnon bands  are given by
 \begin{align}
 \varepsilon_{\eta\bo}^F&= h_0+\eta\sqrt{\bold{h}(\bo)\cdot \bold{h}(\bo)},
\end{align}
where $\eta=\pm$ labels the top and the bottom  bands respectively.  Henceforth we set $S=1/2$. The Floquet magnon bands are depicted in Fig.~\ref{band} for zeroth order $(i)$ and first order $(ii)$ contributions to the effective Hamiltonian.  In the former $(i)$, the system exhibits Dirac magnon points, but they are different  from the Dirac magnon points in the undriven systems \cite{frans} because the Floquet Dirac magnons can be manipulated by anisotropic laser field amplitude, {\it i.e.}, $A_0=(A_x,A_y, 0)$. In this case the Dirac magnon points can move away from the high symmetry points of the Brillouin zone at $\pm{\bf K}=(\pm4\pi/3\sqrt{3},0)$ and can also be gapped out by fine-tuning the anisotropic  amplitudes. However, the system will still remain topologically trivial at the zeroth order as time-reversal symmetry is preserved.

 A nontrivial band topology arises by lowering the frequency and the first order correction becomes important, which breaks time-reversal symmetry. This leads to a synthetic DMI which induces a gap ($\Delta_{\text{gap}}\sim 2|D_F|$) at the Dirac points as shown in Fig.~\ref{band}($ii$). The Floquet magnon bands now acquire a nonzero Berry curvature which has the  form $\Omega_{\eta \bo}^F=\lb\boldsymbol{\nabla}\times \boldsymbol{\mathcal A}_{\eta\bo}^F\rb_z$ where $\boldsymbol{\mathcal A}_{\eta\bo}^F=i\braket{\psi_{\eta\bo}^F|\boldsymbol{\nabla}|\psi_{\eta\bo}^F}$ is the Berry connection  and $\psi_{\eta\bo}^F$ are the Floquet eigenvectors of $\mathcal H_{\text{eff}}(\bo)$.  The Floquet Chern numbers are given by the integration of the Berry curvatures over the Brillouin zone,
\begin{figure}
\includegraphics[width=3.2in]{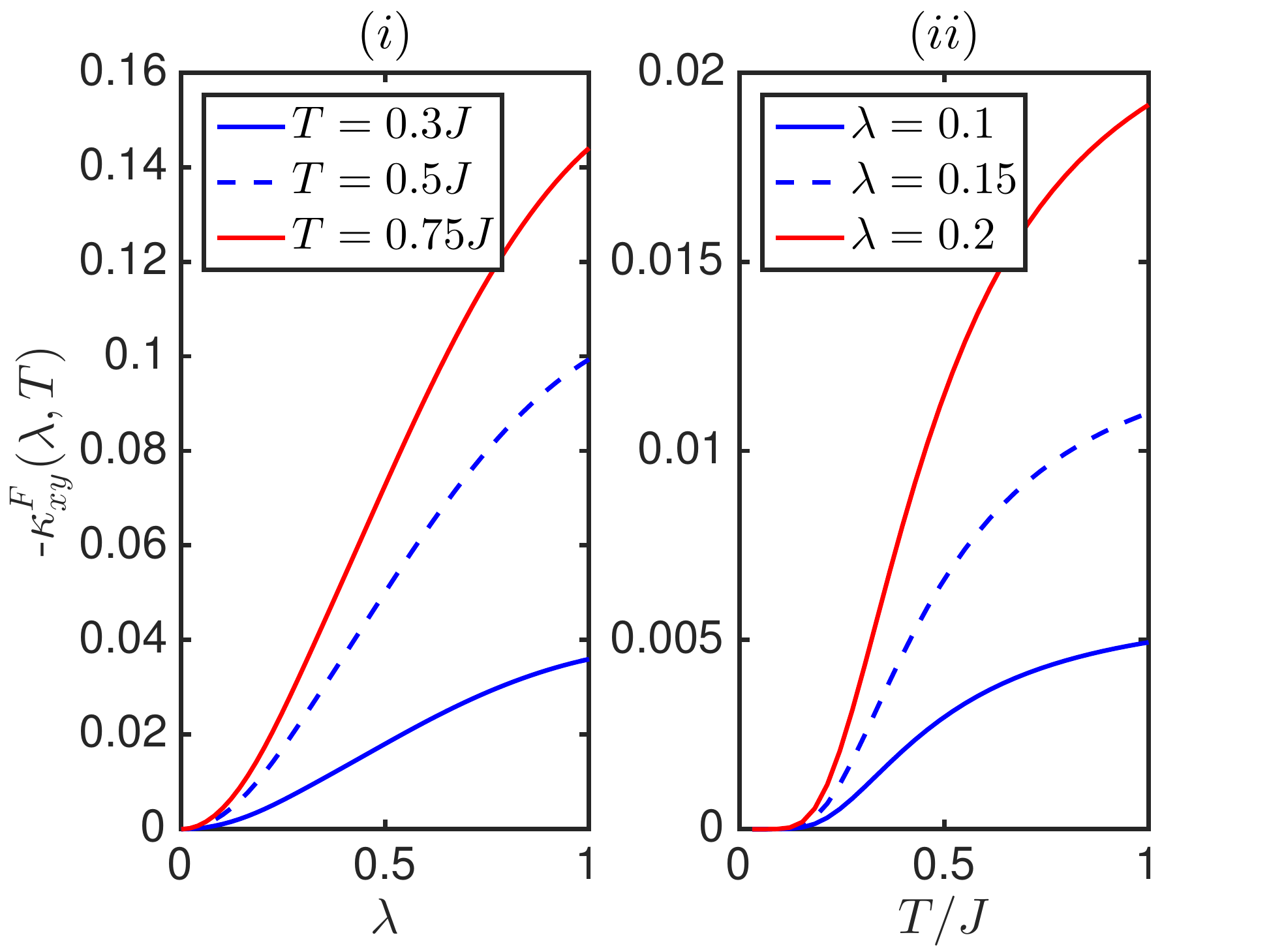}
\caption{Floquet tunable thermal Hall conductivity. $(i)$ $\kappa_{xy}^F$ vs. $\lambda$ for different temperatures. $(ii)$ $\kappa_{xy}^F$ vs. $T$ for different amplitudes. Here,  $J=1$~meV and $\hbar\omega=2.5J$.}
\label{the}
\end{figure}
 \begin{equation}
\mathcal{C}_\eta^F= \frac{1}{2\pi}\int_{{BZ}} d^2k~ \Omega_{\eta\bo}^F.
\label{chenn}
\end{equation}
 It is easily shown that the Floquet Chern numbers are proportional to the synthetic DMI:   $\mathcal{C}_\eta^F\sim \eta \text{sgn}[D_F]$, which can be controlled by the laser field.

 The nontrivial  topology of the Floquet magnon bands has an important transport consequences. It can lead to Floquet thermal magnon Hall effect. This refers to the generation of a  transverse heat  current upon the application of a longitudinal temperature gradient.  In this case, the synthetic laser-field-induced Berry curvature due to the DMI appear in the equations of motion of a magnon wave packet in the same mathematical structure as a magnetic field in the Lorentz force. In other words, the Berry curvature due to the DMI acts like an effective magnetic field in momentum space on the magnons.  The resulting effect is the production of a transverse thermal Hall conductivity, which can derived from linear response theory \cite{alex2}. The Floquet thermal Hall conductivity  is given by 
 \begin{align}
\kappa_{xy}^F(\lambda, T)=-\frac{k_B^2 T}{(2\pi)^2\hbar}\sum_{\eta=\pm}\int_{{BZ}} d^2k c_2\lb n_\eta\rb\Omega_{\eta\bo}^F,
\label{thm}
\end{align}
where
$n_\eta\equiv n_B(\varepsilon_{\eta\bo})=(e^{{\varepsilon_{\eta\bo}}/k_BT}-1)^{-1}$ is the Bose function close to thermal equilibrium, $c_2(x)=(1+x)\lb \ln \frac{1+x}{x}\rb^2-(\ln x)^2-2\textrm{Li}_2(-x),$ and $\text{Li}_n(x)$ is a polylogarithm.  
In  Fig.~\ref{the} we have shown the plots of $\kappa_{xy}^F(\lambda, T)$ as functions of the parameters in units of $k_B/\hbar$. For the two-band honeycomb ferromagnets the Floquet thermal Hall conductivity is negative and it is tunable by the amplitude (frequency) of the laser field. In particular, $\kappa_{xy}^F(\lambda, T)$ can be tuned off at the zeros of $\mathcal J_n(\lambda)$.  This is usually not possible in materials with a strong  intrinsic DMI  in the absence of a laser field.

\begin{figure}
\includegraphics[width=4in]{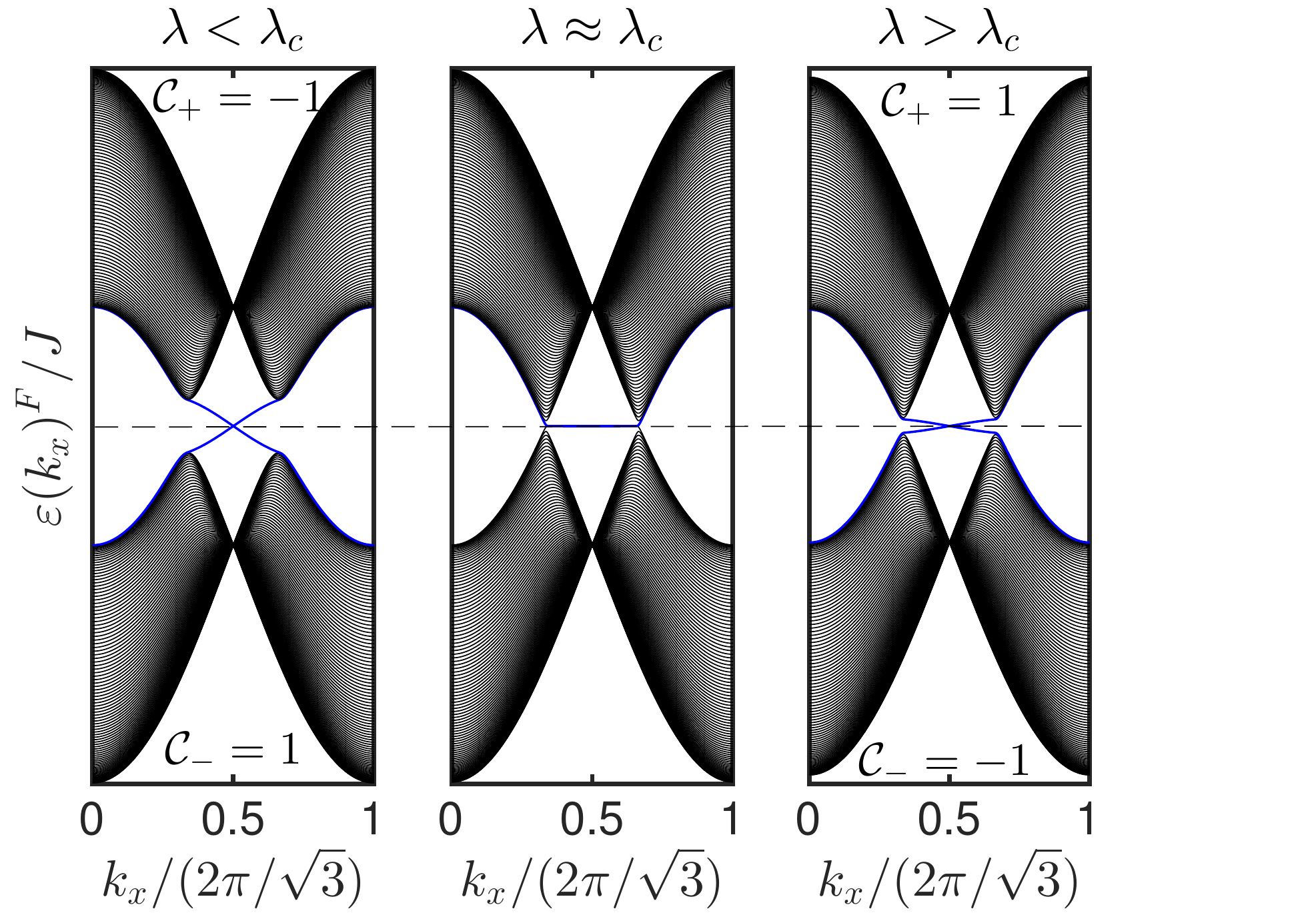}
\caption{Evolution of sign change in the Floquet Chern numbers of honeycomb ferromagnets with tunable intrinsic DMI for different values of $\lambda$ in units of $g\mu_Ba/\hbar c^2$ and $D_0/J=0.15$. The black lines are the magnon bulk bands and blue lines denote the chiral magnon edge modes gapless at  $\varepsilon(k_x=\pi/\sqrt{3})=h_0$ as indicated by the  dashed line.}
\label{edge}
\end{figure}

However, applying circularly polarized laser field in magnetic  materials with a strong  intrinsic DMI can induce a tunable  DMI. Using the simple honeycomb ferromagnetic model as an example, the suitable DMI due to inversion symmetry breaking of the lattice is of the form \cite{sol} $ \mathcal H_{DMI}=D_0\sum_{\la \la \alpha\beta\ra\ra} \nu_{\alpha\beta}\bold{\hat z}\cdot{\bf S}_{\alpha}\times{\bf S}_{\beta}$, where the summation is taken over the triangular plaquettes of the NNN sites.  In the presence of a circularly polarized laser field with sufficiently high frequency, the zeroth order correction to the effective Hamiltonian gives $ \mathcal H_{DMI}^{\text{eff}}=D_0\mathcal J_0(\lambda)\sum_{\la \la \alpha\beta\ra\ra} \nu_{\alpha\beta}\bold{\hat z}\cdot{\bf S}_{\alpha}\times{\bf S}_{\beta}$. Now the value of the intrinsic DMI can be manipulated  by varying $ \lambda$. For $\lambda<\lambda_c \approx 2.4048$ a gap exists at the Dirac points and it closes near the first zero of the Bessel function $\lambda\approx\lambda_c$ and reopens for $\lambda>\lambda_c$. Consequently, the sign of the Chern numbers (Berry curvatures) of the two bands  changes as shown in Fig.~\ref{edge}. Hence, a sign change emerges in  $\kappa_{xy}^F(\lambda, T)$ (not shown). We note that a sign change is not possible  in the undriven system on the honeycomb lattice \cite{sol1,kkim}. The complete topological phase transition including a trivial insulating regime with zero Chern number can be found by including NNN Heisenberg interaction as we have shown in the supplemental material \cite{sm}.  

It is important to note that the circularly polarized  laser field is necessary to induce a tunable synthetic DMI in quantum ferromagnets with negligible   intrinsic DMI. However, in frustrated  magnets with a coplanar magnetic order, a static magnetic field applied perpendicular to the plane of the magnets can induce a tunable synthetic scalar spin  chirality due to non-coplanar spin configurations. In this case  tunable topological magnons can be produced even in the absence of an intrinsic DMI \cite{sol2,sol3,sol4}.

 In conclusion, we have shown that the time-dependent Aharonov-Casher phase acquired by charge-neutral magnons in a  time varying circularly polarized  electric (laser) field induces a synthetic DMI, leading to nontrivial topological magnons in the absence of an intrinsic DMI. Therefore, Dirac magnons and nodal-line magnons become magnon Chern insulators and Weyl magnons respectively under radiation. Our results also showed that the intrinsic DMI in insulating quantum magnets without an inversion symmetry can be manipulated  with circularly polarized  laser field. Hence, the emergent thermal Hall effect can be turned. It is experimentally feasible to investigate the proposed phenomena in magnetic insulators using ultrafast terahertz spectroscopy \cite{pri}.   This formalism may find application in the switching of magnetization and magnon spin current by a laser field as an important step towards magnon spintronics and magnetic data storage \cite{tak4,tak4a}. We believe that these results will extend the study of topological magnons to different magnetic materials without any restrictions.  In an inhomogeneous static electric field  a time-independent Aharonov-Casher phase can be acquired by magnons and leads to magnonic Landau levels in insulating magnets \cite{kouki}.

Research at Perimeter Institute is supported by the Government of Canada through Industry Canada and by the Province of Ontario through the Ministry of Research
and Innovation.

\end{document}

\title{Floquet Topological Magnons: Supplemental Material}
\author{S. A. Owerre}
\affiliation{Perimeter Institute for Theoretical Physics, 31 Caroline St. N., Waterloo, Ontario N2L 2Y5, Canada.}

\maketitle

\section{Topological magnon phase diagram}
 For a fixed $\lambda$ such that $\mathcal J_n(\lambda)\neq 0$  the  honeycomb Floquet topological magnons maps to  the static XXZ ferromagnetic quantum spin Hamiltonian with DM interaction
\begin{align}
\mathcal H&=  -\sum_{\la \alpha,\beta\ra }[J_\perp({S}_{\alpha}^x{S}_{\beta}^x+{S}_{\alpha}^y{S}_{\beta}^y) + J_z{S}_{\alpha}^z{S}_{\beta}^z]\label{model}\\&\nonumber +D\sum_{\la \la \alpha\beta\ra\ra} \nu_{\alpha\beta}\bold{\hat z}\cdot{\bf S}_{\alpha}\times{\bf S}_{\beta}-J_2\sum_{\la\la \alpha,\beta\ra\ra }{\bf S}_{\alpha}\cdot{\bf S}_{\beta}.
\end{align}
We have added an isotropic next-nearest-neighbour (NNN) ferromagnetic exchange coupling $J_2$ and $J_z> J_\perp$ as before. In the bosonic representation we obtain \begin{align}
\mathcal H&=-t_\perp\sum_{\la \alpha\beta \ra}( a_{\alpha}^\dagger a_{\beta}+ h.c.)  - t_2\sum_{\la \la \alpha\beta\ra\ra}(e^{i\phi_{\alpha\beta}}a_{\alpha}^\dagger a_{\beta}+ h.c.)\label{hop}\\&\nonumber +t_A\sum_{\alpha\in A} a_{\alpha}^\dagger a_{\alpha}+t_B\sum_{\alpha\in B} a_{\alpha}^\dagger a_{\alpha}.
\end{align}
The mean-field energy has been dropped. $t_\perp=J_\perp S$, $t_2=\sqrt{t_2^{\prime 2} +t_D^2}$, $t_2^{\prime}(t_D)=J_2S(DS)$, $t_{A(B)}=3J_{z,A(B)}S +6 J_2S$.  The phase factor is $\phi_{\alpha\beta}=\nu_{\alpha\beta}\phi$, where $\phi=\arctan(t_D/t_2^\prime)$ is a magnetic flux generated by the DM interaction on the NNN sites. 

Next, we Fourier transform the magnon hopping Hamiltonian \eqref{hop} into momentum space and the resulting Hamiltonian can be written as $\mathcal{H}=\sum_{\bo}\psi^\dg_\bo\cdot \mathcal{H}(\bo)\cdot\psi_\bo $, with $\psi^\dg_\bo= (a_{\bo A}^{\dg},\thinspace a_{\bo B}^{\dg})$. The  momentum space Hamiltonian is given by
\begin{align}
\mathcal{H}(\bo)=d_0(\bo){\bf I} +\bold{d}(\bo)\cdot\boldsymbol{\sigma},
\end{align}
where  {\bf I}  is an identity $2\times 2$ matrix and $\boldsymbol{\sigma}=(\sigma_x, \sigma_y, \sigma_z)$ are Pauli matrices and $\bold{d}=(d_x, d_y, d_z)$ are given by
\begin{align}
d_0(\bo)&=\epsilon_0-2t_2\cos\phi\sum_j\cos \bo\cdot{\bf b}_j,\\
d_x(\bo)&=-t_\perp\sum_j\cos \bo\cdot\boldsymbol{\delta}_j,\\
d_y(\bo)&=-t_\perp\sum_j\sin \bo\cdot\boldsymbol{\delta}_j,\\
d_z(\bo)&=M-2t_2\sin\phi\sum_j\sin \bo\cdot{\bf b}_j,
\end{align}
with $\epsilon_0= (t_{A}+t_{B})/2 + 6J_2 S$ and $M= (t_{A}-t_{B})/2$. The magnon bands are given by
 \begin{align}
 \varepsilon_s(\bo)=d_0 \pm \sqrt{{\bf d}\cdot {\bf d}}=d_0 + s \epsilon(\bo),
 \end{align}
 where $s=\pm$ for the upper and lower magnon bands respectively.  The  the eigenvectors  are given by
\begin{figure}
\includegraphics[width=3in]{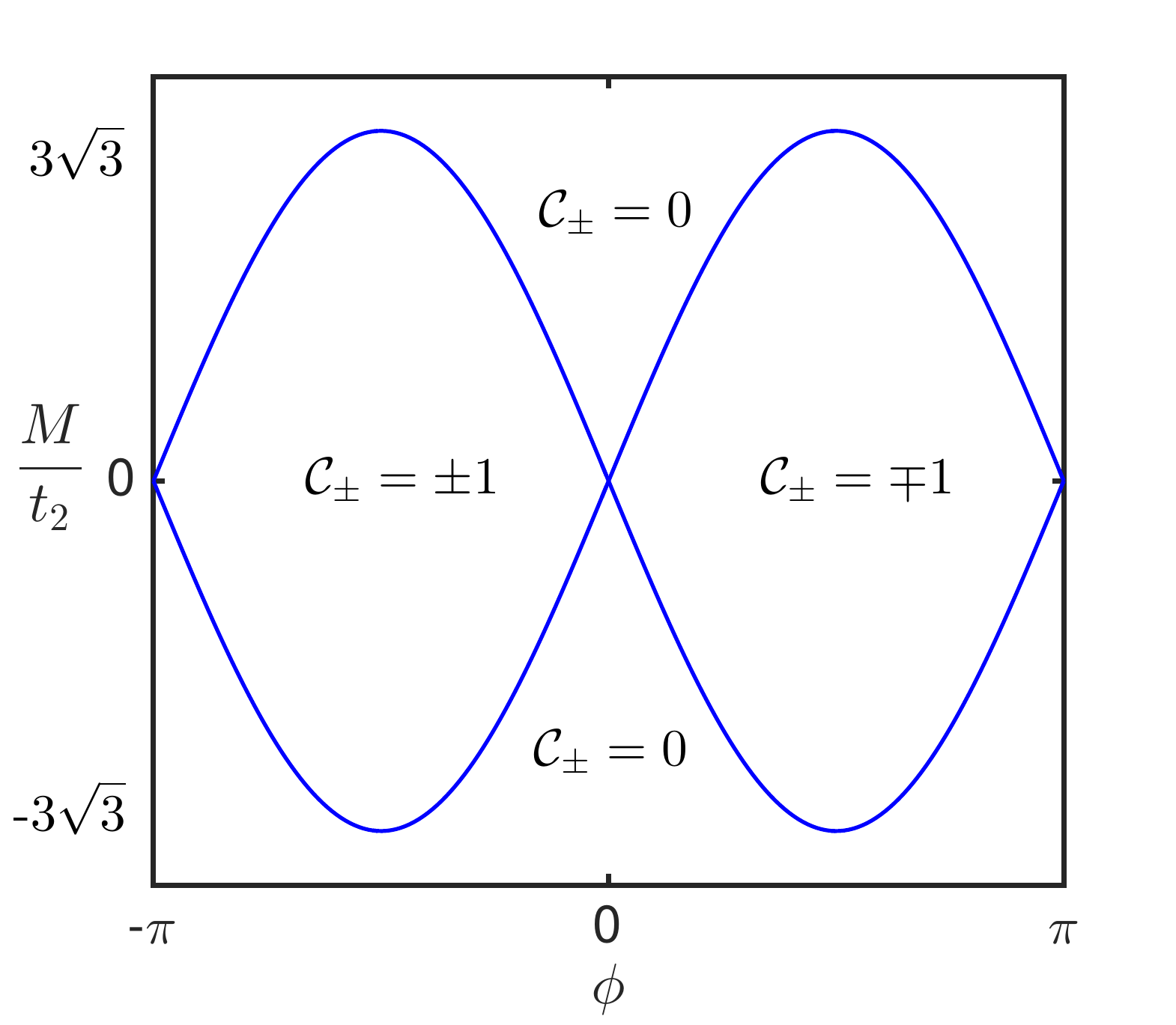}
\caption{The complete phase diagram of the Haldane magnon insulator. In the topological magnon insulator regime $\mathcal C_{\pm}=\pm 1$ and in the trivial magnon insulator   $\mathcal C_{\pm}=0$.}
\label{phase}
\end{figure}

\begin{equation} \ket{\psi_{s\bo}} = \frac{1}{\sqrt{2}}\lb\sqrt{1+s\frac{d_z(\bold k)}{\epsilon(\bold k)}},~ -s e^{-i\varphi_{\bf k}}\sqrt{1-s\frac{d_z(\bold k)}{\epsilon(\bold k)}}\rb^T\label{egv},\end{equation} where 
 \begin{equation}\tan\varphi_{\bf k}= \frac{d_y(\bold k)}{d_x(\bold k)}.\end{equation}
 
 \subsection{Magnon phase transition}
\begin{figure}
\includegraphics[width=3.5in]{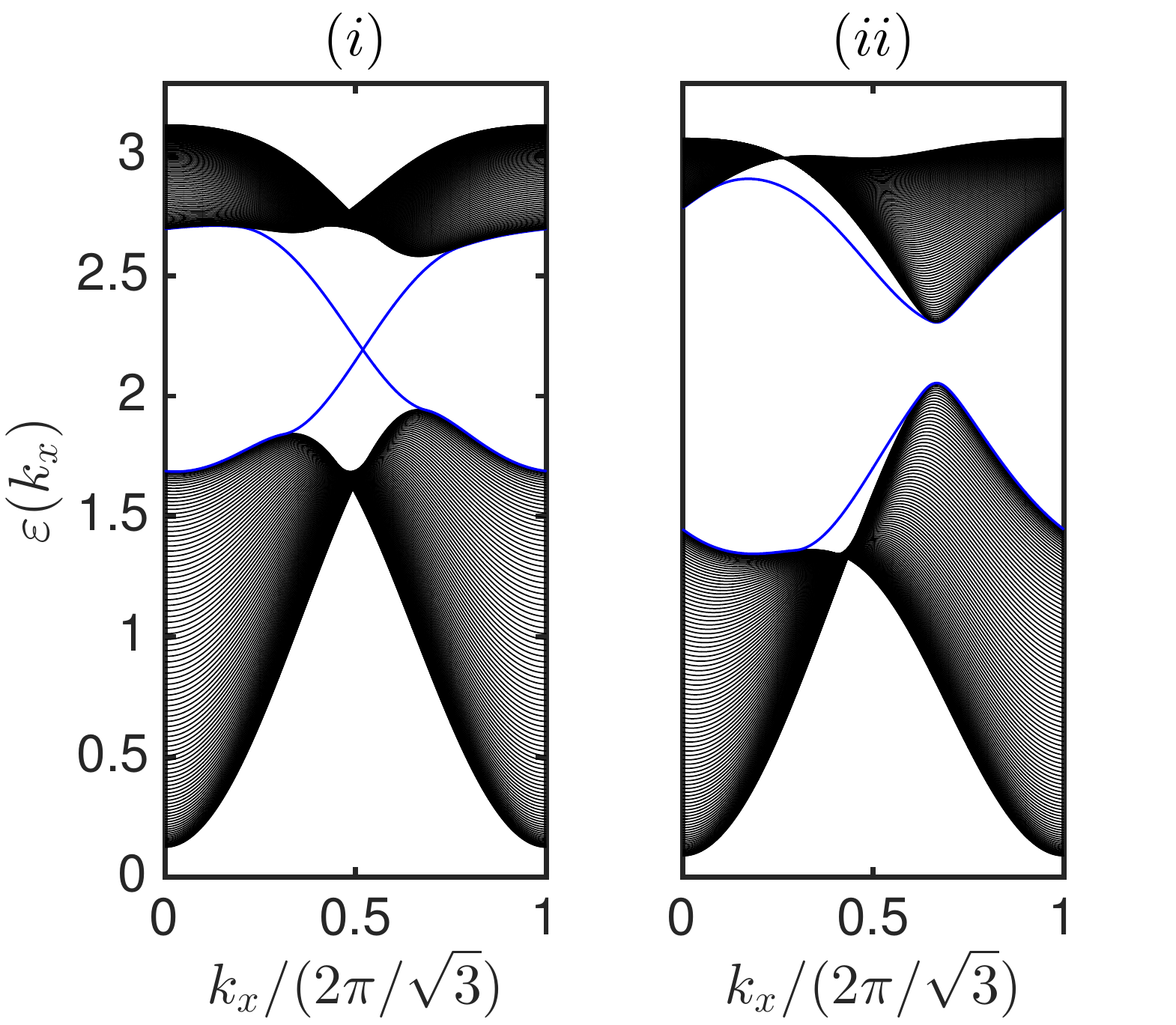}
\caption{Topological and trivial magnon insulator phases on the honeycomb quantum ferromagnets for $(t_2^\prime=t_D=t_2=0.2t_\perp)$ or $\phi=\pi/4$. The parameters  are chosen such that the phase transition occurs at $(M/t_\perp)_c\approx 0.73$ where the magnon bulk gap closes (not shown). $(i)$ Topological Chern magnon insulator with gapless magnon edge modes (blue lines) for $M/t_\perp=.1$. $(ii)$  Trivial magnon insulator with bulk gap and no gapless magnon edge modes (blue lines) for $M/t_\perp=1$. }
\label{edge}
\end{figure}

  The model above is  termed a Haldane magnon insulator. It has only been studied in the topological phase, which is the only phase one would get at the isotropic point $J_\perp=J_z$ \cite{sol,sol1}(see also \cite{kkim}). Therefore, the full topological phase diagram of the Haldane magnon insulator has not been mapped out. In this section, we will present the  complete topological phase diagram of Haldane magnon insulator and also show that it differs from the electronic counterpart \cite{fdm} in terms of transport properties. For $\phi\neq n\pi$, the  magnon bands have  nonzero Berry curvature $\Omega_{s\bo}=\lb\boldsymbol{\nabla}\times \boldsymbol{\mathcal A}_{s\bo}\rb_z$ where $\boldsymbol{\mathcal A}_{s\bo}=i\braket{\psi_{s\bo}|\boldsymbol{\nabla}|\psi_{s\bo}}$ is the Berry connection.  The Chern number is given by

\begin{equation}
 \mathcal{C}_{s}=\frac{1}{2\pi}\int d^2k\Omega_{s\bo}.
\label{thou1}
\end{equation}
The maximum contribution to the Berry curvature comes from the states near the corners of the Brillouin zone ${\bf K}_\tau=(\tau 4\pi/3\sqrt{3},0)$ with $\tau=\pm$. Therefore, we  expand the Hamiltonian  near this point and the   differential operator can be expressed in polar coordinates ($k$,$\phi_k$) where $k=|\bo|$. The Berry curvature takes the form
\bea
\Omega_{ks}^\tau= \frac{1}{k}\lb\frac{\partial{(k\mathcal A_{s\varphi_k}^\tau)}}{\partial k}-\frac{\partial \mathcal A_{s k}^\tau}{\partial\varphi_k}\rb,
\label{becu}
\eea
 where  $\mathcal A_{s \varphi_{k}}^\tau=\frac{1}{k}i\braket{\psi_{ks}^\tau|\partial_{\varphi_k}\psi_{s k}^\tau}$ and $\mathcal A_{s k}^\tau=i\braket{\psi_{ks}^\tau|\partial_{k}\psi_{ks}^\tau}$. We note that $\varphi_{k}$ appears as a U(1) phase in the eigenvectors, therefore $\mathcal A_{s\varphi_k}^\tau$ and $\mathcal A_{s k}^\tau$ are independent of $\varphi_{k}$, thus the second term in Eq.~\ref{becu} vanishes and the resulting expression becomes
\begin{equation}
\Omega_{ks}^\tau= \frac{1}{k}\frac{\partial}{\partial k}\mathcal{P}_{s}^\tau(k); \quad \mathcal{P}_{s}^\tau(k)=i\braket{\psi_{s k}^\tau|\partial_{\phi_k}\psi_{s k}^\tau}.
\label{thou2}
\end{equation} 
By integrating the Berry curvature we obtain 
\begin{align}
\mathcal C_s&=\sum_{\tau=\pm}[\mathcal{P}_{s}^\tau(\infty)-\mathcal{P}_{s}^\tau(0)],\\&=-\frac{s}{2}\sum_{\tau=\pm}\tau\text{sgn}[M +3\sqrt{3}\tau t_2\sin\phi].
\end{align}
A topological magnon phase transition occurs at $M_c=3\sqrt{3}t_2\sin\phi$. The regime $M<M_c$ corresponds to a topological magnon Chern  insulator with $\mathcal C_\pm=\mp 1$ and the regime $M>M_c$ corresponds to a trivial magnon insulator with $\mathcal C_{\pm}=0$.The phase diagram of the Haldane magnon insulator is depicted in Fig.~\ref{phase}. From bulk-edge correspondence we expect  chiral magnon edge modes in the vicinity of the bulk gap, which are protected by the Chern numbers.  As shown in  Fig.~\ref{edge}, the topological regime possesses gapless chiral magnon edge modes whereas the chiral magnon edge modes in the trivial magnon insulator regime contribute to the bulk magnon bands. 

\subsection{Magnon quantum anomalous Hall effect}
 In electronic systems, quantum anomalous Hall effect refers to the generation of a transverse quantized Hall conductivity without an applied magnetic field. Basically, an electric field $\boldsymbol {\mathcal E}$  induces a charge current $\boldsymbol {\mathcal J}_e$ and the Berry curvature acts as an effective magnetic field in momentum space. For insulating quantum ferromagnets the magnon excitations are charge-neutral and do not experience  a Lorentz force. Instead,  a temperature gradient $-\boldsymbol { \nabla} { T}$ induces a heat current $\boldsymbol {\mathcal J}_Q$, and the Berry curvature due to the DMI appears in the equations of motion of a magnon wave packet in the same mathematical structure as a magnetic field in the Lorentz force. In ferromagnets, this does not require an applied  magnetic field analogous to the electronic counterparts. The thermal Hall conductivity is given by  \begin{align}
\kappa_{xy}=-\frac{k_B^2 T}{(2\pi)^2}\sum_{s=\pm}\int_{{BZ}} d^2k c_2\lb n_s\rb\Omega_{s\bo},
\label{thm}
\end{align}
where
$n_s\equiv n_B[\varepsilon_{s\bo}]=[e^{{\varepsilon_{s\bo}}/k_BT}-1]^{-1}$ is the Bose function, $c_2(x)=(1+x)\lb \ln \frac{1+x}{x}\rb^2-(\ln x)^2-2\textrm{Li}_2(-x),$ and $\text{Li}_n(x)$ is a polylogarithm.  
 
The major difference here is that magnon transports require a finite temperature due to the Bose function.  In other words, magnonic systems do not have conduction and valence bands and no bands are fully occupied or completely empty. Instead, the magnons can be thermally populated at low temperatures near the $\bo=0$ mode. In fact, the maximum contribution of $\kappa_{xy}$ comes from the lowest magnon band near the  $\bo=0$ mode at low temperatures.

 Unlike electronic systems, the topological and the trivial insulating regimes in Fig.~\ref{edge} will possess a finite  $\kappa_{xy}$ due to the population of the band near the $\bo=0$ mode at low temperatures. In this limit the Berry curvature of the lowest band in the topological and trivial insulating phases has the form \begin{align}
\Omega_-(\bold k\to 0)\simeq \frac{c_1 k_xk_y^2 +c_2k_x^2k_y^2}{c_3},
\end{align}
where $c_1= -27(M t_2)/8$, $c_2= 81\sqrt{3}t_2\sin\phi/8$ and $c_3=4(M^2 + 9t_2^2)^{3/2}$.  In the low-temperature limit $\kappa_{xy}$ can be well-approximated as 
 \begin{align}
\kappa_{xy} &\simeq \frac{1}{T}\int \frac{d^2k}{(2\pi)^2} n_B[\varepsilon_-(\bold k)][\varepsilon_-(\bold k)+\varepsilon_+(\bold k)]^2\Omega_-(\bold k).
\label{thm4}
\end{align}
Near the $\bold k =0$ mode, $\varepsilon_-(\bold k)+\varepsilon_+(\bold k)\simeq 2(t_A+t_B)=t_0$ and   $\varepsilon_-(\bold k)\simeq a+ bk^2$, where $a=(t_A+t_B)/2 - 3t_\perp$ and $b=3t_\perp/4$. Now, we can parameterize the momentum in the polar coordinate $(k_x=k\cos\theta,~k_y=k\sin\theta)$ and perform the angular part of the integration,  
 \begin{align}
\kappa_{xy} &\simeq\frac{c_2}{c_3}\frac{t_0^2}{16\pi T}\int_0^\infty k ~dk \frac{k^4}{e^{(a +b k^2)/T}-1},\nonumber\\&=\frac{T^2}{16\pi b^3}\frac{c_2}{c_3}\text{Li}_{3}\bigg[ \exp{\lb-\frac{a}{T}\rb}\bigg].
\label{thm5}
\end{align}
 This is the low-temperature thermal Hall conductivity in the topological and trivial insulating phases.